\documentclass[aps,prd,preprint,floatfix,nofootinbib,groupedaddress,showpacs]{revtex4}
\usepackage{graphicx}
\usepackage{amsmath}
\usepackage{latexsym}
\usepackage{here}
\usepackage{array}
\usepackage{dcolumn}
\usepackage{bm}
\usepackage{epsfig}

\newcommand{\dd}{\mbox{{\rm d}}}

\newcommand{\Lumint}{{\cal L}_{\rm int}}

\begin{document}



\title{Revealing the large extra dimension effective interaction
at an $e^+e^-$ collider with polarized beams}


\author{A. A. Pankov}
\email[]{pankov@ictp.it}
\author{A. V. Tsytrinov}
\email[]{atsytrin@ictp.it}
\affiliation{ICTP Affiliated Centre,
Pavel Sukhoi Technical University, Gomel 246746, Belarus}
\author{N. Paver}
\email[]{nello.paver@ts.infn.it}
\affiliation{University of Trieste and INFN-Trieste Section, 34100
Trieste, Italy}
\date{\today}
\begin{abstract}

%
%
%

Several types of New Physics scenarios are represented by
contactlike effective interactions. An example is the exchange of
nonstandard quanta of very large mass scales, beyond the
kinematical limit for direct production set by the available
collider energy. This kind of interactions can be revealed only
through deviations of observables from the Standard Model
predictions. If such deviations were observed, the relevant source
should be identified among the possible models that could explain
them. Here, we assess the expected ``identification reach'' on the
ADD model of gravity in large compactified extra dimensions,
against the compositeness-inspired four-fermion contact
interaction. As basic observables we take the differential cross
sections for fermion-pair production at a 0.5-1 TeV
electron-positron linear collider with both beams longitudinally
polarized. For the four-fermion contact interaction we assume a
general linear combination of the individual models with definite
chiralities, with arbitrary coupling constants. In this sense, the
estimated identification reach on the ADD model can be considered
as `model-independent'. In the analysis, we give estimates also
for the expected ``discovery reaches'' on the various scenarios.
We emphasize the substantial r\^ole of beams polarization in
enhancing the sensitivity to the contactlike interactions under
consideration.

\end{abstract}


\maketitle
\section{Introduction}
Numerous New Physics (NP) scenarios are described by local, contactlike,
effective interactions between the Standard Model (SM) particles.
This is the typical case of interactions mediated by exchanges of
quanta that are constrained, by either conceptual or phenomenological
considerations, to have a mass $\Lambda$ in the multi-TeV range. These
states may be beyond the kinematical reach of the collider and therefore could
not appear as final products of the studied reactions. Accordingly,
the existence of such nonstandard scenarios can be verified
only through their {\it indirect} effects, represented by deviations of
the measured observables from the SM predictions. The effective
interaction theoretical framework leads to the expansion of the deviations 
caused by these novel interactions in powers of the corresponding
small ratios $E_{\rm C.M.}/\Lambda\ll 1$, multiplied
by matrix elements of local operators between initial and final
states. Generally, the dominance of the leading power is taken as a
reasonable assumption.
\par
Referring to experiments at planned high energy colliders and their
sensitivity to NP, one can define for the individual contactlike 
effective interactions the expected {\it discovery reach}, as the maximum
value of the relevant $\Lambda$ for which deviations from the SM
predictions can be detected within the foreseen experimental accuracy. This
limit can be assessed by a comparison of theoretical deviations, functions of
$\Lambda$, and expected experimental uncertainties {\it via} a $\chi^2$
procedure, by assuming that no such deviations are observed.
\par
Conversely, one can envisage a situation where corrections to the SM
predictions are observed, and found compatible with one of the effective
interactions for a certain value of the relevant $\Lambda$. In this case,
one should consider that, in principle, different contactlike
interactions can cause similar corrections. Therefore, it should be desirable
to attempt the identification of the source of the observed deviations
among the various possible scenarios. To this purpose, one can define the
expected {\it identification reach} on an individual
contact interaction model, as the maximum value of the
corresponding $\Lambda$ for which not only it can cause observable
deviations but, also, can be discriminated as the source of such corrections,
were they observed, against the other effective interactions for any value
of their characteristic $\Lambda$s. Obviously, the identification reach
can only be smaller than the discovery reach.
\par
Here, we consider as basic observables the differential cross
sections for the fermion pair production processes
\begin{equation}
e^++e^-\to{\bar f}+f,\qquad\quad f=e,\mu,\tau,c,b, \,
\label{proc}
\end{equation}
at the International Linear Collider (ILC) with longitudinally polarized
electron and positron beams \cite{Aguilar-Saavedra:2001rg}. This option is
considered with great interest in the project for this collider, and its
impact on the physics programme has been reviewed recently in
Ref.~\cite{Moortgat-Pick:2005cw}.
\par
As a significant example, we focus on the identification
reach on the ADD model of gravity in large, compactified, extra
spatial dimensions
\cite{Arkani-Hamed:1998rs,Arkani-Hamed:1998nn,Antoniadis:1998ig},
with respect to the compositeness-inspired four-fermion contact
interactions \cite{Eichten:1983hw,Ruckl:1983hz}. In particular,
we insist on the r\^ole played by the longitudinal polarization of the
$e^+$ and $e^-$ beams in enhancing the identification power of
processes (\ref{proc}) on this scenario, at the planned ILC energies
and luminosities.
\par
Specifically, in Sects.~2 and 3 we collect the explicit expressions of
the polarized differential distributions and, to fix notations,
briefly introduce the contactlike interactions of interest here and the
corresponding corrections to the SM amplitudes; in Sect.~4, after deriving
the foreseeable discovery reaches on the individual models, we present an
assessment of the model-independent identification reach on the ADD
model against the four-fermion contact interaction (the specific meaning
of `model-independence' being clarified below); finally, Sect.~5 is
devoted to some conclusive remarks.
\section{Polarized differential distributions}
Neglecting all fermion masses with respect to the c.m. energy
$\sqrt s$, the expression of the polarized differential cross section for the
process $e^+e^-\to f\bar{f}$ with $f\ne e,t$ can be expressed as
\cite{Schrempp:1987zy,Pankov:2005ar}:
\begin{equation}
\frac{d\sigma(P^-,P^+)}{d z}=\frac{D}{4} \left[(1-P_{\rm eff})
\left(\frac{d\sigma_{\rm LL}}{d z} +\frac{d\sigma_{\rm LR}}{d
z}\right) +(1+P_{\rm eff})\left(\frac{d\sigma_{\rm RR}}{d z}
+\frac{d\sigma_{\rm RL}}{d z}\right)\right]. \label{crossdif-pol}
\end{equation}
In Eq.~(\ref{crossdif-pol}), $z=\cos\theta$ with $\theta$ the angle between
initial and final fermions in the C.M. frame, and the
subscripts $\rm L$, $\rm R$
denote the respective helicities. Furthermore, with $P^-$ and $P^+$ denoting
the degrees of longitudinal polarization of the $e^-$ and $e^+$ beams,
respectively, one has
\cite{Flottmann:1995ga,Fujii:1995ys}
\begin{equation}
D=1-P^- P^+\hskip 2pt , \qquad P_{\rm eff}=\frac{P^--P^+}{1-P^-
P^+}. \label{poleff}
\end{equation}
The SM amplitudes for these processes are determined by $\gamma$ and $Z$
exchanges in the $s$-channel.
\par
The polarized differential cross section for the Bhabha process
$e^+e^-\to e^+e^-$, where $\gamma$ and $Z$ can be exchanged also in the
$t$-channel, can be conveniently written as
\cite{Pankov:2002qk,Pankov:2005kd}:
\begin{eqnarray}
\frac{\dd\sigma(P^-,P^+)}{\dd z}
&=&\frac{(1+P^-)\,(1-P^+)}4\,\frac{\dd\sigma_{\rm R}}{\dd z}+
\frac{(1-P^-)\,(1+P^+)}4\,\frac{\dd\sigma_{\rm L}}{\dd z}
\nonumber \\
&+&\frac{(1+P^-)\,(1+P^+)}4\,\frac{\dd\sigma_{{\rm RL},t}}{\dd z}+
\frac{(1-P^-)\,(1-P^+)}4 \,\frac{\dd\sigma_{{\rm LR},t}}{\dd z},
\label{cross}
\end{eqnarray}
with the decomposition
\begin{eqnarray}
\frac{\dd\sigma_{\rm L}}{\dd z}= \frac{\dd\sigma_{{\rm LL}}}{\dd z
}+ \frac{\dd\sigma_{{\rm LR},s}}{\dd z}, \qquad
\frac{\dd\sigma_{\rm R}}{\dd z}= \frac{\dd\sigma_{{\rm RR}}}{\dd
z}+ \frac{\dd\sigma_{{\rm RL},s}}{\dd z}. \label{sigP}
\end{eqnarray}
In Eqs.~(\ref{cross}) and (\ref{sigP}), the subscripts $t$ and $s$ denote
helicity cross sections with SM $\gamma$ and $Z$ exchanges in the
corresponding channels. In terms of helicity amplitudes:
\begin{eqnarray}
\frac{\dd\sigma_{{\rm LL}}}{\dd z}&=&
\frac{2\pi\alpha_{\rm e.m.}^2}{s}\,\big\vert G_{{\rm LL},s}+
G_{{\rm LL},t} \big\vert^2, \quad\ \ \frac{\dd\sigma_{{\rm
RR}}}{\dd z}= \frac{2\pi\alpha_{\rm e.m.}^2}{s}\,\big\vert
G_{{\rm RR},s}+G_{{\rm RR},t}\big\vert^2 ,
\nonumber \\
\frac{\dd\sigma_{{\rm
LR},t}}{\dd z}&=&\frac{\dd\sigma_{{\rm
RL},t}}{\dd z}= \frac{2\pi\alpha_{\rm
e.m.}^2}{s}\,\big\vert G_{{\rm LR},t}\big\vert^2,
\quad\frac{\dd\sigma_{{\rm LR},s}}{\dd z}=
\frac{\dd\sigma_{{\rm RL},s}}{\dd z}=
\frac{2\pi\alpha_{\rm e.m.}^2}{s}\,\big\vert G_{{\rm
LR},s}\big\vert^2. \label{helsig}
\end{eqnarray}
\par
The polarized differential cross section (\ref{crossdif-pol}) for the
leptonic channels $e^+e^-\to l^+l^-$ with $l=\mu,\tau$
can be obtained directly from Eq.~(\ref{cross}), basically by dropping
the $t$-channel poles. The same is true, after some obvious adjustments, for
the ${\bar c}c$ and ${\bar b}b$ final states.
\par
According to the previous considerations the amplitudes
$G_{\alpha\beta,i}$, with
$\alpha,\beta={\rm L,R}$ and $i=s,t$, are given by the sum of the SM
$\gamma, Z$ exchanges plus deviations representing the effect of the novel,
contactlike, effective interactions:
\begin{eqnarray}
G_{{\rm LL},s}&=&{u}\,\left(\frac{1}{s}+\frac{g_{\rm L}^2
}{s-M^2_Z}+ \Delta_{{\rm LL},s}\right), \quad G_{{\rm
LL},t}={u}\,\left(\frac{1}{t}+\frac{g_{\rm L}^2}{t-M^2_Z}+
\Delta_{{\rm LL},t}\right),
\nonumber \\
G_{{\rm RR},s}&=&{u}\,\left(\frac{1}{s}+ \frac{g_{\rm
R}^2}{s-M^2_Z}+ \Delta_{{\rm RR},s}\right), \quad G_{{\rm
RR},t}={u}\,\left(\frac{1}{t}+\frac{g_{\rm R}^2}{t-M^2_Z}+
\Delta_{{\rm RR},t}\right),
\nonumber \\
G_{{\rm LR},s}&=&{t}\,\left(\frac{1}{s}+\frac{g_{\rm R}\hskip 2pt
g_{\rm L}}{s-M^2_Z}+\Delta_{{\rm LR},s}\right), \qquad G_{{\rm
LR},t}= s\,\left(\frac{1}{t}+\frac{g_{\rm R}\hskip 2pt g_{\rm
L}}{t-M^2_Z}+\Delta_{{\rm LR},t}\right). \label{helamp}
\end{eqnarray}
Here $u,t=-s(1\pm z)/2$, $g_{\rm R}=\tan\theta_W$ and
$g_{\rm L}=-\cot{2\,\theta_W}$ with $\theta_W$ the electroweak mixing angle.
The deviations $\Delta_{\alpha\beta,i}$ caused by the models of interest here
have been tabulated in earlier references, see for example
Refs.~\cite{Pankov:2005kd,Pasztor:2001hc,Cullen:2000ef}. However, for
convenience, we report their explicit expressions and briefly comment on their
properties in the next section.
\section{Effective interactions and deviations from the SM}
The contactlike nonstandard interactions considered in the sequel are
listed below:
\par
{\bf a)} The ADD, compactified large extra dimensions, scenario
\cite{Arkani-Hamed:1998rs,Arkani-Hamed:1998nn,Antoniadis:1998ig},
motivated by the gauge hierarchy problem. In this scenario, only
gravity can propagate in the full multidimensional space.
Correspondingly, a tower of graviton KK states with equally-spaced
spectrum is exchanged in the ordinary four-dimensional space, and
induces indirect corrections to the SM $\gamma$ and $Z$
exchanges. The relevant Feynman rules have been derived in
Refs.~\cite{Han:1998sg,Giudice:1998ck}. In the parameterization of
Ref.~\cite{Hewett:1998sn}, the exchange of such a KK tower is
represented by the effective interaction:
\begin{equation}
{\cal L}=i\frac{4\lambda}{\Lambda_H^4}T^{\mu\nu}T_{\mu\nu},
\qquad \lambda=\pm 1.
\label{dim-8}
\end{equation}
In Eq.~(\ref{dim-8}), $T_{\mu\nu}$ denotes the energy-momentum tensor
of the SM particles and $\Lambda_H$ is an ultraviolet cut-off on the
summation over the KK spectrum, expected in the (multi) TeV range.
The corresponding corrections to the SM amplitudes for Bhabha scattering, 
see Eq.~(\ref{helamp}), read:
\begin{equation}
\Delta_{{\rm LL},s} = \Delta_{{\rm RR},s}=
\frac{\lambda}{\pi\alpha_{\rm e.m.}\Lambda_H^4} (u+\frac{3}{4}s),\qquad
\Delta_{{\rm LL},t}=\Delta_{{\rm RR},t}=
\frac{\lambda}{\pi\alpha_{\rm e.m.}\Lambda_H^4} (u+\frac{3}{4}t),
\nonumber
\end{equation}
\begin{equation}
\Delta_{{\rm LR},s} = -\frac{\lambda}{\pi\alpha_{\rm e.m.}\Lambda_H^4}
(t+\frac{3}{4}s), \qquad\ \ \ \ \ \ \
\Delta_{{\rm LR},t}=-\frac{\lambda}{\pi\alpha_{\rm e.m.}\Lambda_H^4}
(s+\frac{3}{4}t).
\label{deltadd}
\end{equation}
As observed in the previous section, the deviations for the other
processes in Eq.~(\ref{proc}) can easily be obtained from
Eqs.~(\ref{deltadd}). One can remark, also, that the effective interaction
(\ref{dim-8}) has dimension-8, which explains the high negative power of
the characteristic mass scale $\Lambda_H$.
\par
{\bf b)} The dimension-6 four-fermion contact interaction (CI)
scenario \cite{Eichten:1983hw,Ruckl:1983hz}. With $\Lambda_{\alpha\beta}$
($\alpha,\beta={\rm L,R}$) the ``compositeness'' mass scales,
and $\delta_{ef}=$1 (0) for $f=e$ ($f\ne e$):
\begin{equation}
{\cal L}=\frac{4\pi}{1+\delta_{ef}}\hskip 3pt \sum_{\alpha,\beta}
\hskip 3pt
\frac{\eta_{\alpha\beta}}{\Lambda^2_{\alpha\beta}} \left(\bar
e_\alpha\gamma_\mu e_\alpha\right) \left(\bar f_\beta\gamma^\mu
f_\beta\right), \qquad \eta_{\alpha\beta}=\pm 1,0.
\label{CI}
\end{equation}
The induced deviations in Eq.~(\ref{helamp}) are:
\begin{equation}
\Delta_{{\alpha\beta},s}=\Delta_{{\alpha\beta},t}=
\frac{1}{\alpha_{\rm e.m.}}\frac{\eta_{\alpha\beta}}{\Lambda_{\alpha\beta}^2}.
\label{deltaci}
\end{equation}
Rather generally, this kind of effective interactions applies to the cases
of very massive virtual exchanges, such as heavy $Z^\prime$s, leptoquarks,
{\it etc}.
\par
Current experimental lower bounds on $\Lambda$s are mostly derived
from nonobservation of deviations at LEP and Tevatron
colliders. At the 95\% C.L., they are: $\Lambda_H>1.3\, {\rm TeV}$
\cite{Cheung:2004ab} and, generically, $\Lambda_{\alpha\beta}> 10-15\,
{\rm TeV}$, depending on the processes measured and the type of analysis
performed \cite{Yao:2006px}.
\par
It may be worth noticing that in case {\bf b)}, Eq.~(\ref{deltaci}),
the deviations are $z$-independent and the appropriate helicity
cross sections have the same angular structure as in the case of
the SM. Conversely, in case {\bf a)}, Eq.~(\ref{deltadd}), the deviations
introduce extra $z$-dependencies in the angular distributions. In turns out
that, as a consequence, the ADD model contribution to the integrated cross
sections for the annihilation channels in Eq.~(\ref{proc}) is quite
 small, due to the vanishing interference with the SM amplitudes after
integration over the full angular range $-1\le z\le 1$. This suppresses
the possibility of identifying the ADD interaction effects in the total
cross sections for these processes. In these cases, specifically defined
integrated asymmetries with polarized initial beams may be expected to
be more efficient contactlike interaction analyzers
\cite{Pankov:2005ar,Osland:2003fn}. In the
next section we discuss the r\^ole of polarized angular differential
distributions themselves, in selecting signatures of ADD effective
interactions at ILC.
\section{Discovery and identification reaches from polarized differential
cross sections}
An approach based on the polarized differential distributions for
the lepton-pair production processes in (\ref{proc}) was proposed in
Ref.~\cite{Pankov:2005kd}. Here, {\it pairs} of individual 
effective contactlike interactions, (\ref{dim-8}) and (\ref{CI}) with
currents of definite chirality, were compared to each other as sources of
deviations from the SM predictions,
and estimates of the foreseeable discrimination reaches from each other at ILC
were obtained by a simple $\chi^2$ analysis. In the sequel, we extend the
analysis of \cite{Pankov:2005kd} in two respects:
\par
{\it i}) The discovery and identification potential on $\Lambda_H$ of
quark-pair production processes in Eq.~(\ref{proc}) is considered.
\par
{\it ii}) We take into account the fact that, for a given flavour
of the fermionic final state $f$, Eq.~(\ref{CI}) defines four independent
CI models. In principle, then, one can
envisage the more general possibility that the four-fermion contact
interaction is a linear combination of the individual four-fermion
operators in Eq.~(\ref{CI}) with free, simultaneously nonvanishing,
independent coupling constants $\eta_{\alpha\beta}/\Lambda_{\alpha\beta}^2$.
Accordingly, we estimate the foreseeable identification reach
on the ADD mass scale $\Lambda_H$ in the case where
the corrections to the SM amplitudes simultaneously depend
on {\it all} independent mass scales $\Lambda_{\alpha\beta}$ in
Eq.~(\ref{CI}), in addition to $\Lambda_H$.
In the analysis of ref.~\cite{Pankov:2005kd}, only one CI coupling
at a time had been separately assumed as a free, potentially nonzero,
parameter.
The situation considered here is therefore much more general and, in this
sense, the corresponding estimate of the identification reach on
$\Lambda_H$ should be considered as `model-independent'.
\par
To derive the constraints on the models, one has to compare the
theoretical deviations from the SM predictions, that are functions of
$\Lambda$s, to the foreseen experimental uncertainties on the
differential cross sections. To this purpose, taking the polarized angular
distributions as basic observables for the analysis,
${\cal O}={\rm d}\sigma(P^-,P^+)/{\rm d} z$, we
introduce the relative deviations from the SM predictions and the
corresponding $\chi^2$:
\begin{equation}
\Delta ({\cal O})=\frac{{\cal O}(\rm SM+NP)-{\cal O}(\rm
SM)}{{\cal O} (\rm SM)};\qquad \chi^2({\cal O})=\sum_{\{P^-,\ P^+\}}
\sum_{\rm bins}\left(\frac{\Delta({\cal O})^{\rm bin}}
{\delta{\cal O}^{\rm bin}}\right)^2. \label{reldev}
\end{equation}
Here, for the individual processes, the cross sections for the
different initial polarization configurations are combined in the $\chi^2$,
and $\delta{\cal O}$ denotes the expected experimental
relative uncertainty (statistical plus systematic one).
As indicated in Eq.~(\ref{reldev}), we divide the
angular range into bins. For Bhabha scattering, the cut angular range
$\vert\cos\theta\vert< 0.90$ is divided into ten equal-size bins. Similarly,
for annihilation into muon, tau and quark pairs we consider the analogous
binning of the cut angular range $\vert\cos\theta\vert< 0.98$.
\par
For the Bhabha process, we combine the cross sections with the
following initial electron and positron longitudinal polarizations:
$(P^-,P^+)=(\vert P^-\vert,-\vert P^+\vert)$;
$(-\vert P^-\vert,\vert P^+\vert$;
$(\vert P^-\vert,\vert P^+\vert)$; $(-\vert P^-\vert,-\vert P^+\vert)$.
For the ``annihilation'' processes in
Eq.~(\ref{proc}), with $f\ne e,t$, we limit to combining the
$(P^-,P^+)=(\vert P^-\vert,-\vert P^+\vert)$ and
$(-\vert P^-\vert,\vert P^+\vert)$ polarization configurations.
Numerically, we take the ``standard'' envisaged values $\vert P^-\vert=0.8$
and $\vert P^+\vert=0.6$.
\par
Regarding the ILC energy and time-integrated luminosity, for
simplicity we assume the latter to be equally distributed among
the different polarization configurations defined above. The
explicit numerical results will refer to C.M. energy $\sqrt s=0.5$
TeV with time-integrated luminosity ${\cal L}_{\rm int}= 100$
$fb^{-1}$, and to $\sqrt s=1$ TeV with ${\cal L}_{\rm int}=1000$
$fb^{-1}$. The assumed reconstruction efficiencies, that determine
the expected statistical uncertainties, are 100\% for $e^+e^-$
final pairs; 95\% for final $l^+l^-$ events ($l=\mu,\tau$); 35\%
and 60\% for $c {\bar c}$ and $b {\bar b}$, respectively. The
major systematic uncertainties are found to originate from
uncertainties on beams polarizations and on the time-integrated
luminosity: we assume $\delta P^-/P^-=\delta P^+/P^+=0.2$\% and
$\delta{\cal L}_{\rm int}/{\cal L}_{\rm int}=0.5$\%, respectively.
\par
As theoretical inputs, for the SM amplitudes we use the
effective Born approximation \cite{Consoli:1989pc}
with $m_{\rm top}=175~\text{GeV}$ and $m_{\rm H} = 120~\text{GeV}$.
Concerning the ${\cal O}(\alpha)$ QED corrections, the (numerically dominant)
effects from initial-state radiation for Bhabha scattering and the
annihilation processes in (\ref{proc}) are accounted for by a structure
function approach including both hard and soft photon emission
\cite{nicrosini}, and by a flux factor method \cite{physicsatlep2},
respectively. Effects of radiative flux return to the $s$-channel $Z$
exchange are minimized by the cut
$\Delta\equiv E_\gamma/E_{\rm beam}< 1-M_Z^2/s$ on the radiated
photon energy, with $\Delta= 0.9$. In this way, only interactions that occur
close to the nominal collider energy are included in the analysis and,
accordingly, the sensitivity to the manifestations of the searched for
nonstandard physics can be optimized. By a calculation based on the
ZFITTER code \cite{Bardin:1999yd}, other QED
effects such as final-state and initial-final state emission are
found, in processes $e^+e^-\to l^+l^-$ and $e^+e^-\to\bar{q}q$
($q=c,b$), to be numerically unimportant for the chosen kinematical
cuts. Finally, correlations between the different polarized cross
sections (but not between the individual angular bins) are taken into
account in the derivation of the numerical results presented in the
next subsections.
\subsection{Discovery reaches}
The expected discovery reaches on the contactlike effective
interactions are assessed by assuming a situation where no
deviation from the SM predictions is observed within the
experimental uncertainty. Accordingly, the corresponding upper
limits on the accessible values of $\Lambda$s are determined by
the condition $\chi^2({\cal O})\le\chi^2_{\rm CL}$, and we take
$\chi^2_{\rm CL}=3.84$ for a 95\% C.L.
\par
In Table~\ref{table:DIS}, we present the numerical results from
the processes listed in the caption,\footnote{Here, $l^+l^-$
denotes the combination of $\mu^+\mu^-$ and $\tau^+\tau^-$ final states,
and $\mu-\tau$ universality has been assumed for the limits on CI mass
scales.} at an ILC with $\sqrt s=0.5$ TeV, ${\cal L}_{\rm int}=100$ $fb^{-1}$, 
and with $\sqrt s=1$ TeV, ${\cal L}_{\rm int}=1000$ $fb^{-1}$. 
In this table, only the results for
positive interference between SM amplitudes and nonstandard
contributions are reported, i.e., the cases $\lambda=1$ for the ADD model of
Eq.~(\ref{dim-8}) and $\eta_{\alpha\beta}=1$ for the CI models
of Eq.~(\ref{CI}). Indeed, the sensitivity reach for negative interference
turns out to be practically the same. The results
in Table~\ref{table:DIS} clearly show the enhancement in sensitivity to
the considered effective interactions allowed, for given C.M. energy and
luminosity, by beams polarization. This effect is particularly substantial
in the case of the CI models (\ref{CI}), for which the limits on the
relevant $\Lambda$s are quite high compared to the current ones.
\begin{table}[!htb]
\caption{95\% C.L. discovery reaches (in TeV). Left and right
entries in each column refer to the polarizations $(\vert
P^-\vert,\vert P^+\vert$)=(0,0) and (0.8,0.6), respectively. }
\begin{tabular}{|l|c|c|c|c|} \hline
\raisebox{-1.50ex}[0cm][0cm]{Model}&
\multicolumn{4}{c|}{Process} \\
 &
$e^{+}e^{-} \to e^{+}e^{-}$ & $e^{+}e^{-} \to l^{+}l^{-}$
& $e^{+}e^{-} \to \bar{b}b$ & $e^{+}e^{-} \to \bar{c}c$ \\ \hline 
 & \multicolumn{4}{c|}{ $\sqrt{s} = 0.5$ TeV, $\Lumint = 100fb^{-1}$} \\
\cline{2-5}
$ \Lambda_{H} $ & 4.1; 4.3& 3.0; 3.2& 3.0; 3.4& 3.0; 3.2 \\
$\Lambda_{VV}^{ef}$& 76.2; 86.4& 89.7; 99.4& 76.1; 96.4& 84.0; 94.1 \\
$\Lambda_{AA}^{ef}$& 47.4; 69.1& 80.1; 88.9& 76.7; 98.2& 76.5; 85.9 \\
$\Lambda_{LL}^{ef}$& 37.3; 52.5& 53.4; 68.3& 63.6; 72.7& 54.5; 66.1 \\
$\Lambda_{RR}^{ef}$& 36.0; 52.2& 51.3; 68.3& 42.5; 71.2& 46.3; 66.8 \\
$\Lambda_{LR}^{ef}$& 59.3; 69.1& 48.5; 62.8& 51.3; 68.7& 37.0; 57.7 \\
$\Lambda_{RL}^{ef}$& $\Lambda_{RL}^{ee} = \Lambda_{LR}^{ee}$
& 48.7; 63.6& 46.8; 60.1& 52.2; 60.7 \\
\cline{2-5} 
 & \multicolumn{4}{c|}{ $\sqrt{s} = 1$ TeV, $\Lumint = 1000fb^{-1}$} \\
\cline{2-5}
$ \Lambda _{H} $ & 8.7; 9.4& 6.7; 7.0& 6.7; 7.5& 6.7; 7.1 \\
$\Lambda _{VV}^{ef}$& 173.6; 205.1& 218.8; 244.3& 185.6; 238.2&
206.2;
232.3 \\
$\Lambda _{AA}^{ef}$& 109.9; 166.1& 194.7; 217.9& 186.; 242.7&
186.4;
210.8 \\
$\Lambda _{LL}^{ef}$& 83.7; 122.8& 128.3; 165.5& 154.5; 175.8&
131.3;
159.6 \\
$\Lambda _{RR}^{ef}$& 80.5; 122.1& 123.4; 166.1& 103.5; 176.9&
111.8;
164.1 \\
$\Lambda _{LR}^{ef}$& 136.6; 166.8& 120.5; 156.6& 124.9; 170.2&
92.7;
144.6 \\
$\Lambda _{RL}^{ef}$& $\Lambda _{RL}^{ee} = \Lambda _{LR}^{ee}$ &
120.8; 158.3& 120.1; 151.9& 129.6; 151.1 \\ \hline
\end{tabular}
\label{table:DIS}
\end{table}
\subsection{Identification reach on the ADD scenario}

Continuing the previous $\chi^2$-based analysis, we now assume
that deviations has been observed and are consistent with the ADD
scenario (\ref{dim-8}) for some value of $\Lambda_H$. To assess the
level at which the ADD model can be discriminated from the general CI
model as the source of the deviations or, equivalently, to determine
the `model-independent' identification reach on the effective interaction 
({\ref{dim-8}), we introduce in analogy with Eq.~(\ref{reldev}) 
the relative deviations ${\tilde\Delta}$ and the corresponding 
${\tilde\chi}^2$:
\begin{equation}
{\tilde\Delta} ({\cal O})= \frac{{\cal O}({\rm CI})-{\cal O}({\rm
ADD})}{{\cal O}({\rm ADD})}; \qquad\quad {\tilde\chi}^2({\cal O})=
\sum_{\{P^-,\ P^+\}}\sum_{\rm bins}\left (\frac{{\tilde\Delta}({\cal
O})^{\rm bin}} {{\tilde\delta}{\cal O}^{\rm bin}}\right)^2.
\label{chitilde}
\end{equation}
\begin{figure}[!htb]
\vspace*{-1.0cm} \centerline{ \hspace*{-0.3cm}
\includegraphics[width=8.4cm,angle=0]{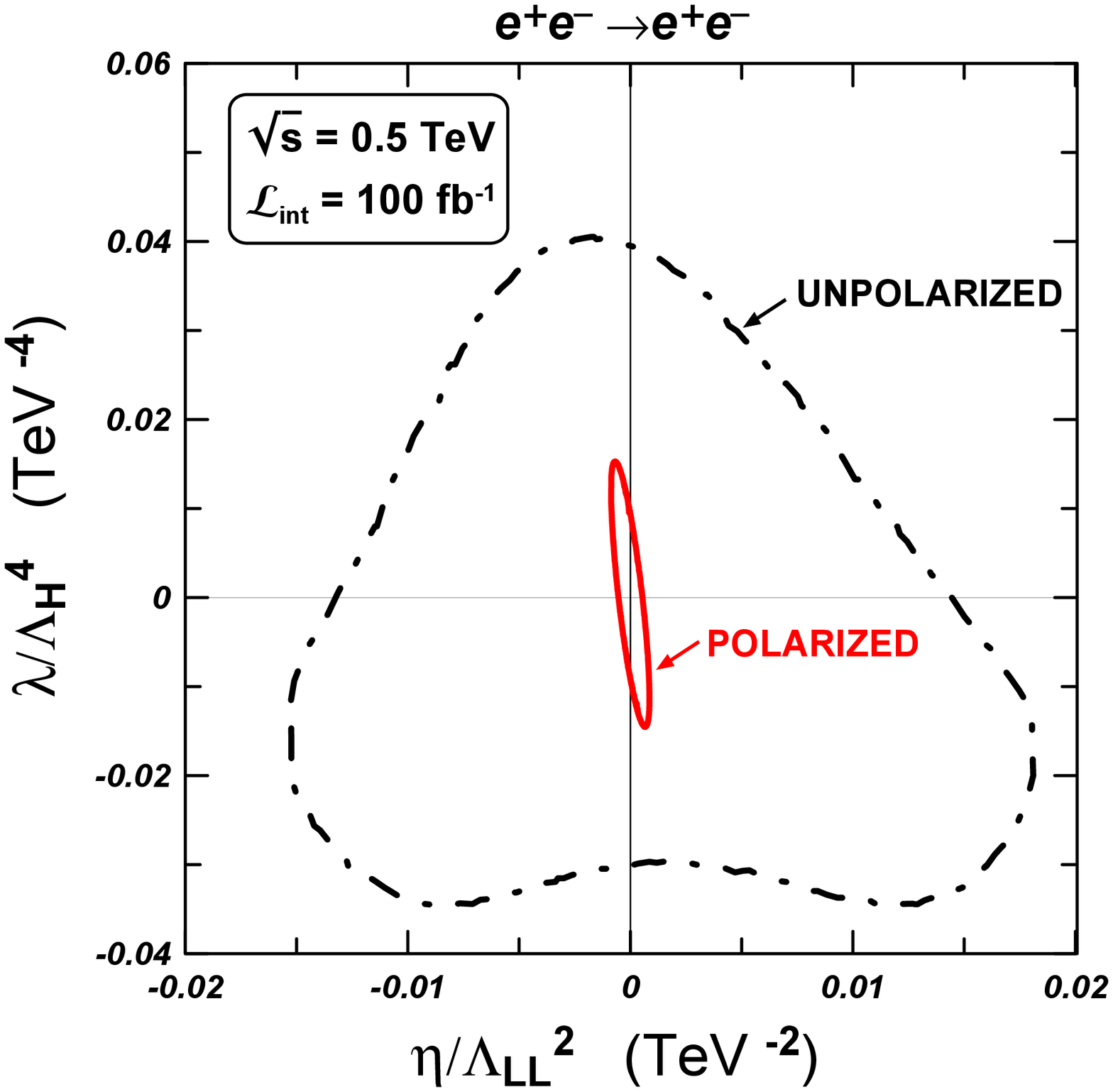}
\hspace*{-0.55cm}
\includegraphics[width=8.4cm,angle=0]{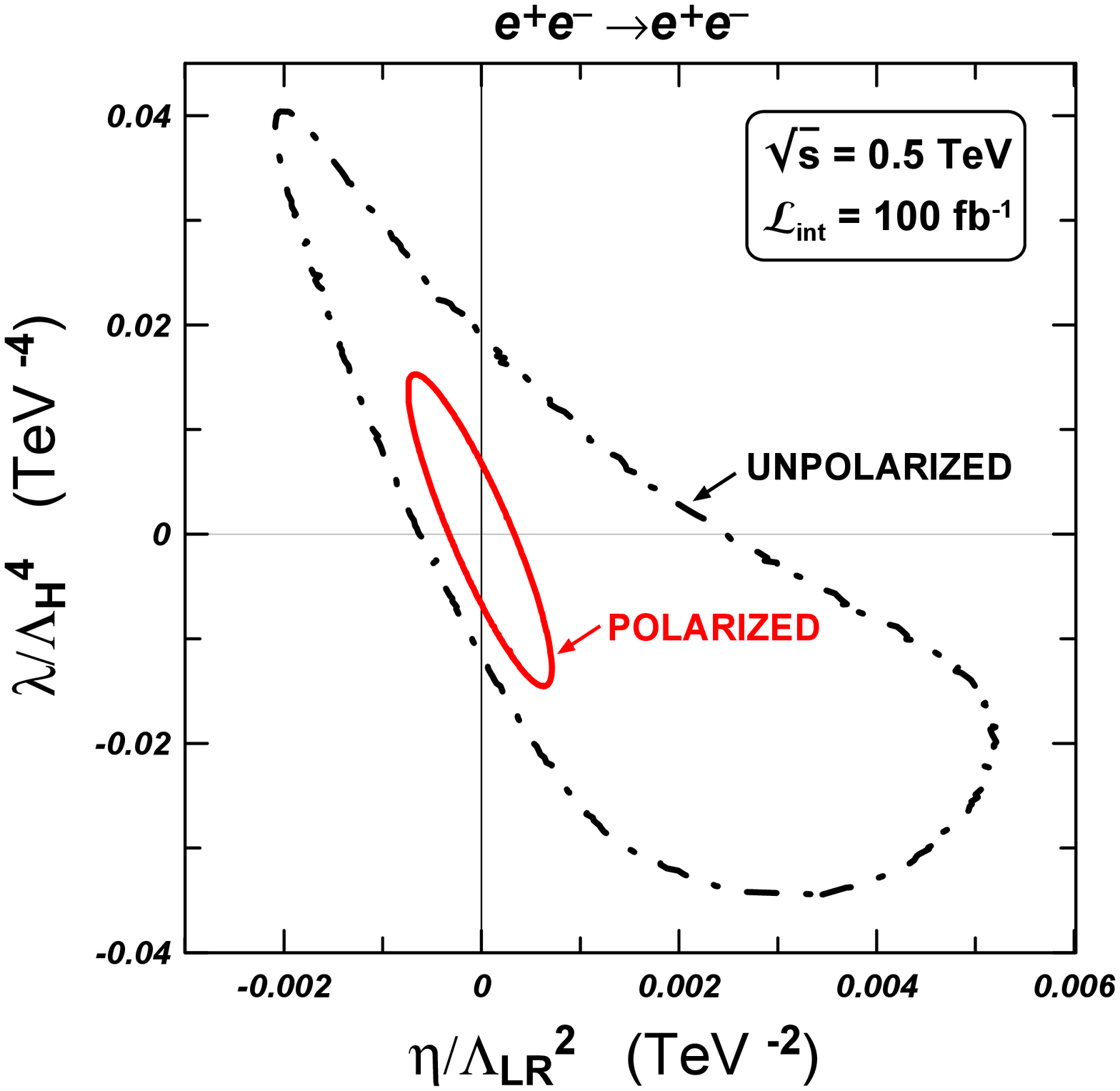}}
\vspace*{-2.0cm} \caption{\label{fig:1} Two-dimensional projection
of the 95\% C.L. confusion region onto the planes
$(\eta_{\rm LL}/\Lambda_{\rm LL}^2$, $\lambda/\Lambda_H^4)$ (left
panel) and $(\eta_{\rm LR}/\Lambda_{\rm LR}^2$ ,
$\lambda/\Lambda_H^4)$ (right panel) obtained from Bhabha
scattering with unpolarized beams (dot-dashed curve) and with both
beams polarized (solid curve). } \label{fig1}
\end{figure}
In Eq.~(\ref{chitilde}), ${\tilde\Delta} ({\cal O})$ depends on
all $\Lambda$s, and somehow represents the `distance' between the
ADD and the CI model in the parameter space ($\Lambda_H,
\Lambda_{\alpha\beta}$). Moreover, ${\tilde\delta{\cal O}^{\rm
bin}}$ is the expected relative uncertainty referred to the cross
sections that include the ADD model contributions: its statistical
component is therefore determined from helicity amplitudes with
the deviations (\ref{deltadd}) predicted for the given value of
$\Lambda_H$. In turn, the CI contributions to the cross sections
bring in the dependence of Eq.~(\ref{chitilde}) on the parameters
$\Lambda_{\alpha\beta}$ of Eq.~(\ref{deltaci}), now considered 
as {\it all} independent.
Therefore, for each of process (\ref{proc}), ${\tilde\chi}^2$ is a
function of $\lambda/\Lambda_H^4$ and in general, as anticipated
in sect.~3, simultaneously of the four CI couplings
$\eta_{\alpha\beta}/(\Lambda^{ef}_{\alpha\beta})^2$.
\par
In this situation we can determine {\it confusion regions} in the parameter
space, where the CI model can be considered as consistent with the ADD model,
in the sense that it can mimic the differential cross sections of the
individual processes (\ref{proc}) determined by the latter one.
At a given C.L., these confusion regions are determined by the condition
\begin{equation}
{\tilde\chi}^2\le\chi^2_{\rm CL}. \label{chi2ID}
\end{equation}
According to the number of independent CI couplings active in the
different processes, for 95\% C.L. we choose $\chi^{2}_{\rm
CL}=7.82$ for Bhabha scattering and $\chi^{2}_{\rm CL}=9.49$ for
lepton ($\mu^+\mu^-$, $\tau^+\tau^-$) and quark ($\bar{c}c$,
$\bar{b}b$) pair production processes.
\par
The simple $\chi^2$ procedure outlined above is clearly `CI
model-independent', and we represent graphically some examples of
the numerical results from Bhabha scattering at $\sqrt{s}=0.5 {\rm
TeV}$ and ${\cal L}_{\rm int}=100 fb^{-1}$. For this process,
Eq.~(\ref{chi2ID}) defines a four-dimensional surface enclosing a
volume in the $(\lambda/\Lambda_H^4$, $ \eta_{\rm LL}/\Lambda_{\rm
LL}^2$, $\eta_{\rm RR}/\Lambda_{\rm RR}^2$, $\eta_{\rm
LR}/\Lambda_{\rm LR}^2)$ parameter space. In Fig.~\ref{fig1}, we
show the planar surfaces that are obtained by projecting the 95\%
C.L. four-dimensional surface, hence the corresponding confusion
region that results from the condition 
${\tilde\chi}^2=\chi^2_{\rm CL}$, onto the
two planes $(\eta_{\rm LL}/\Lambda_{\rm LL}^2$,
$\lambda/\Lambda_H^4)$ and $(\eta_{\rm LR}/\Lambda_{\rm LR}^2$,
$\lambda/\Lambda_H^4)$ (we limit our graphical examples to these
pairs of parameters).
\par
As suggested by Fig.~\ref{fig1}, the contour of the confusion
region turns out to identify a maximal value of
$\vert\lambda/\Lambda_H^4\vert$ (equivalently, a minimum value of
$\Lambda_H$), for which the CI scenario can be excluded at the
95 \% C.L. for any value of $\eta/\Lambda_{\alpha\beta}^2$. This
value, $\Lambda_H^{\rm ID}$, is the identification reach on the
ADD scenario, namely, for $\Lambda_H<\Lambda_H^{\rm ID}$ the CI scenario
can be excluded as explanation of deviations from SM predictions attributed
to the ADD interaction, and the latter can therefore be {\it identified}.
\par Fig.~\ref{fig:1} shows the dramatic r\^ole of initial beams
polarization in obtaining a restricted region of confusion in the
parameter space or, in other words, in enhancing the
identification sensitivity of the differential angular
distributions to $\Lambda_H^{\rm ID}$. Table~\ref{table:IDR} shows
the numerical results for the foreseeable `model-independent'
identification reaches on $\Lambda_H$, for the two choices of C.M.
energy and luminosity made in Table~\ref{table:DIS}.
\begin{table}[!htb]
\caption{95\% CL identification reach on the ADD model parameter
$\Lambda_{H}$ obtained from $e^+e^- \to \bar{f} f$ with
polarizations ($|P^-|$,$|P^+|$)=(0,0) and (0.8, 0.6),
respectively.}
\begin{tabular}{|l|c|c|c|c|}
\hline \raisebox{-1.50ex}[0cm][0cm]{$\Lambda _{H}$ (TeV)}&
\multicolumn{4}{c|}{Process} \\
 & $e^{+}e^{-} \to e^{+}e^{-}$ & $e^{+}e^{-} \to l^{+}l^{-}$ & $e^{+}e^{-} \to \bar{b}b$ & $e^{+}e^{-} \to \bar{c}c$ \\
\hline 
$\sqrt{s} = 0.5$ TeV, $\Lumint = 10^{2}fb^{-1}$ & 2.2; 2.9& 2.3;
2.3& 2.6; 2.9& 2.3; 2.4
\\ \hline 
$\sqrt{s} = 1.0$ TeV, $\Lumint = 10^{3}fb^{-1}$ & 5.0; 6.4& 4.9;
5.1  & 5.8; 6.2& 5.1; 5.3
\\ \hline 
\end{tabular}
\label{table:IDR}
\end{table}
\section{Concluding remarks}
We have presented a simple, $\chi^2$ based, estimate of the power
for searching and distinguishing signatures of spin-2 graviton
exchange envisaged by the ADD model, that is foreseeable at the
polarized ILC with $\sqrt s=0.5$-1 TeV. The basic observables in
the analysis are the polarized differential cross sections for
fermion-pair production processes. The compositeness-inspired
four-fermion contact interaction, from which the ADD model should
be discriminated in case of observation of corrections to the SM
predictions, has been assumed to be of the general form, 
i.e., a linear combination of the individual contact interaction
operators with definite chiralities. The coefficients of such a
combination have been taken into account simultaneously as
independent, and potentially nonvanishing, constants.
\par
The discovery reaches, as well as the identification reaches,
are quite high compared to the current bounds, and depend on
energy and luminosity as shown in Table~\ref{table:DIS} and in
Table~\ref{table:IDR}, respectively. In particular,
Table~\ref{table:IDR} shows that, of the four considered
$e^+e^-$ processes, Bhabha scattering and $\bar{b}b$ pair
production definitely have the best identification sensitivity
on the mass scale $\Lambda_H$ characterizing the ADD model
for gravity in `large' compactified extra dimensions. The
substantial r\^ole of beams polarization is exemplified by
Fig.~\ref{fig1} (where the confusion region between the
considered models is dramatically reduced), and by the
discovery reaches on the models shown in Table~\ref{table:DIS}.
\par
The enhancement of the estimated identification sensitivity on the
ADD effective interaction is quite considerable: as exemplified by
the entries of Table~\ref{table:IDR}, in the polarized case the
identification reach on $\Lambda_H$ ranges from 2.9 TeV to 6.4
TeV, depending on energy, luminosity and degree of longitudinal
polarization. Although unavoidably somewhat depressed by the
penalty due to the general multi-parameter expression assumed for
the CI scenario (that implies taking large values of the
${\chi}^2_{\rm CL}$), these `model-independent' identification
values of $\Lambda_H$ are still much higher than the current
limits. In fact, we find that they are only moderately 
lower (by some 10-20\%) than the `model-dependent' ones 
obtained in Ref.~\cite{Pankov:2005kd} 
by assuming only one nonzero CI coupling at a time.   
These nice features reflect in part the small values
assumed for the relative uncertainties on electron and positron
beams polarization in the previous section, and call for very
high precision on polarimetry measurements at the ILC.

\vspace{0.5cm} \leftline{\bf Acknowledgements}
\par \noindent
AAP acknowledges the support of INFN and of MIUR (Italian Ministry
of University and research). NP has been partially supported by
funds of MIUR and of the University of Trieste. This research has
been partially supported by the Abdus Salam ICTP. AAP and AVT also
acknowledge the Belarusian Republican Foundation for Fundamental
Research.

\end{document}